\documentclass[12pt,preprint]{aastex}

\newcommand{\dmm}{\mbox{$\Delta$m$_{15}(B)$}}
\newcommand{\kms}{\mbox{km s$^{-1}$}}

\shorttitle{SNe 1999cc, 1999cl, and 2000cf}
\shortauthors{Krisciunas et al.}

\begin{document}

\title{Photometry of the Type Ia Supernovae 1999cc, 1999cl, and 2000cf\altaffilmark{1}}

\author{Kevin Krisciunas,\altaffilmark{2}
Jose Luis Prieto,\altaffilmark{3}
Peter M. Garnavich,\altaffilmark{2}
Jessica-Lynn G. Riley,\altaffilmark{4}
Armin Rest,\altaffilmark{5}
Christopher Stubbs,\altaffilmark{6} and
Russet McMillan\altaffilmark{7}
}

\altaffiltext{1}{Based in part on observations at Apache Point
Observatory, which is operated by the Astrophysical Research Corsortium,
and in part on observations with the Vatican Advanced Technology Telecope.}

\altaffiltext{2}{University of Notre Dame, Department of Physics, 225
  Nieuwland Science Hall, Notre Dame, IN 46556-5670;
  {kkrisciu@nd.edu}, {pgarnavi@nd.edu}}

\altaffiltext{3}{Ohio State University, Deptartment of Astronomy,
  4055 McPherson Laboratory, 140 W. 18th Ave., Columbus, Ohio 43210;
  {prieto@astronomy.ohio-state.edu}}

\altaffiltext{4}{Humboldt State University, 1 Harpst St., Arcata, CA 95521;
  {snoflake@hotmail.com}}

\altaffiltext{5}{Cerro Tololo Inter-American Observatory, Casilla
  603, La Serena, Chile; {arest@noao.edu}}

\altaffiltext{6}{Department of Physics and Department of Astronomy,
  17 Oxford Street, Harvard University, Cambridge MA 02138; {cstubbs@fas.harvard.edu}}

\altaffiltext{7}{Apache Point Observatory, Astrophysical Research Consortium,
2001 Apache Point Road, P. O. Box 59, Sunspot, NM 88349-0059; {rmcmillan@apo.nmsu.edu}}

\begin{abstract}
  
We present previously unpublished $BVRI$ photometry of the Type Ia
supernovae 1999cc and 2000cf along with revised photometry of SN~1999cl.
We confirm that SN~1999cl is reddened by highly non-standard dust, with 
R$_V$ = 1.55 $\pm$ 0.08.  Excepting two quasar-lensing galaxies whose
low values of R$_V$ are controversial, this is the only known object with 
a published value of
R$_V$ less than 2.0.  SNe~1999cl and 2000cf have near-infrared absolute magnitudes 
at maximum in good agreement with other Type Ia SNe of mid-range decline rates.  
\end{abstract}
\keywords{supernovae: individual (SN~1999cc, SN~1999cl, SN~2000cf) ---
techniques: photometric}

\section{Introduction}  

Type Ia supernovae (SNe) are the most important extragalactic distance calibrators
beyond a redshift of $z = 0.01$. Since the discovery of the decline rate
relation \citep{Psk77,Psk84,Phi93}, which correlated the absolute magnitudes at
maximum of Type Ia SNe with the rates of decline, a number of methods
have evolved for analyzing the optical light curves: 1) the \dmm\ method
\citep{Ham_etal96, Phi_etal99, Ger_etal04, Pri_etal05};\footnote[8]{\dmm\ was
originally defined to be the number of $B$-band magnitudes that a Type Ia
SN decreases in the first 15 days after maximum light.  The index is now
based on $BVRI$ light curves, and technically \dmm\ is not only a measure
of the $B$-band decline rate, but also implies details about the shape of
the $R$-band shoulder, the shape of the $I$-band secondary hump, and the
relative times of the $BVRI$ maxima.} 2) the multi-color light curve
shape (MLCS) method \citep{Rie_etal96a, Rie_etal98, Jha_etal05b}; 3) the
stretch method, which is applied to $U$-, $B$- and $V$-band light curves
\citep[][and references therein] {Gol_etal01}; and 4) the ``C-magic''
method of \citet{Wan_etal03}.  \citet{Nob_etal03} and \citet{Wan_etal05}
have also devised new uses for optical color indices of Type Ia SNe.  In
this paper we shall rely on the latest version of the \dmm\ method
\citep{Pri_etal05}.

There is a great deal of effort being expended on the discovery and
follow up of high redshift SNe because of the implications for a non-zero
cosmological constant \citep{Rie_etal98, Per_etal99, Ton_etal03,
Kno_etal03, Bar_etal04, Rie_etal04, Kri_etal05}.  The observation and
calibration of nearby Type Ia SNe continues to be important.  Without a
large sample of nearby SNe covering all decline rates and light curve
shapes we cannot make full use of the high redshift data.

Another topic often swept under the rug relates to reddened SNe.  
\citet{Rie_etal96b} give R$_V \equiv$ A$_V /$E($B-V$) = 2.55~$\pm$ 0.30 as
the most appropriate value for a group of 20 Type Ia SNe.  
\citet{Rei_etal05} find R$_V$ = 2.65 $\pm$ 0.15 from a sample of 122 Type Ia
SNe.  But it is often assumed that the dust in other galaxies is exactly
like the average dust in our Galaxy, which has R$_V$ = 3.1
\citep{Sne_etal78, Car_etal89}.

Previously, we found that SN~1999cl must be reddened by very non-standard
dust, with R$_V \approx$ 1.8 \citep{Kri_etal00}. If standard dust
parameters are assumed, SN~1999cl's host (M~88, in the Virgo cluster) is
placed only half-way to the Virgo cluster.  \citet{Eli_etal06} show that
the highly dimmed and reddened SN~2003cg is also affected by host galaxy
dust with R$_V \approx$ 1.8. We know that our Galaxy has dusty lines of
sight that exhibit values of R$_V$ ranging from 2.1 to 5.8 \citep[][and
references therein]{Dra03}.  We should not simply assume that all other
galaxies have dust with R$_V$ = 3.1.

This is the third and last paper of a series that contains optical and
infrared photometry of Type Ia supernovae (SNe) and which was originally
presented in the lead author's Ph.~D. Dissertation \citep{Kri00}.  Previous
papers included data on SNe~1999aa, 1999cl, 1999cp, 1999da, 1999dk, 1999gp,
2000bk, and 2000ce \citep{Kri_etal00, Kri_etal01}.  In this paper we report
previously unpublished $BVRI$ photometry of SNe~1999cc and 2000cf.  These
two SNe were the subjects of an REU\footnote[9]{Research Experiences for
Undergraduates, funded by the National Science Foundation.} project carried
out by one of us (J-LGR).

Our images of SNe~1999cc and 2000cf were obtained with the 3.5-m telescope
at Apache Point Observatory (APO) in New Mexico using its facility CCD
camera Spicam.  Reference images were obtained on 15 March 2004 using the
1.8-m Vatican Advanced Technology Telescope at Mt. Graham, Arizona.  We
also report one $H$-band measurement of SN~2000cf obtained at APO using the
infrared camera Grim II.

Finally, we report revised $BVRI$ photometry of SN~1999cl.  
Some previously published photometry of SN~1999cl \citep{Kri_etal00},
obtained one month after maximum with the 0.9-m reflector of the
University of Washington's Manastash Ridge Observatory (MRO)
required, in retrospect, the use of image subtraction templates,
which were unavailable at that time. Subtraction templates were
obtained at APO on 10 February 2000 UT, some 243 days after
the time of $B$-band maximum of SN~1999cl.  This is sufficiently long
after maximum light that the SN would have dimmed by more than 5
magnitudes.  Thus, systematic errors in the revised photometry
resulting from the possible presence of the SN in the subtraction
templates would be of order 0.01 mag $-$ smaller than the random
errors of the photometry.

Additional $UBVRI$ photometry of these three SNe is given by \citet{Jha02}
and \citet{Jha_etal05a}.  This photometry was obtained at the Fred L.
Whipple Observatory (FLWO) at Mt. Hopkins, Arizona.

\section{Observations}  

SN~1999cc was discovered by \citet{Sch99} on 8.25 May 1999 UT (= JD
2,451,306.75). A spectrum reported by \citet{Gar_etal99} and obtained on
14.3 May UT revealed SN~1999cc to be a Type Ia SN before maximum light.  
This SN was located 16.8 arcsec east and 1.7 arcsec north of the nucleus of
the Sc galaxy NGC 6038.  The radial velocity of the galaxy, corrected to the
frame of the Cosmic Microwave Background (CMB) radiation, is 9452 \kms.

SN~2000cf was discovered by \citet{Puc_Seh00} on 9.23 May 2000 UT (= JD
2,451,673.73). A spectrum obtained by \citet{Jha_etal00} on
11.33 May UT revealed SN~2000cf to be a Type Ia SN several days past maximum
light.  This SN was located 3.1 arcsec east and 4.3 arcsec north of the
nucleus of MCG +11-19-25.  The radial velocity of the galaxy, corrected to
the frame of the Cosmic Microwave Background radiation, is 10803 \kms.  
Thus, both SNe~1999cc and 2000cf are sufficiently distant ($>$ 130 Mpc) that
they partake of the smooth Hubble flow (which begins at a distance of
roughly 40 Mpc).

Figure \ref{99cc_finder} shows the field of SN~1999cc and some nearby field
stars.  The coordinates of the SN and these field stars are to be found in
Table \ref{99cccoords}.  We include the $BVRI$ magnitudes of the field
stars, which were calibrated using observations of \citet{Lan92} standards
on three photometric nights.

A corresponding finder chart for SN~2000cf is to be found in Figure
\ref{00cf_finder}.  Coordinates and $BVRI$ magnitudes of its field stars,
tied to Landolt standards on six photometric nights, are given in 
Table \ref{00cfcoords}.  

We used a combination of software to derive the $BVRI$ magnitudes of SNe
1999cc and 2000cf.  First, we used a package of scripts devised by Brian
Schmidt for the alignment, kernel matching, and subtraction of the images
from the reference templates.  The kernel matching of the Point Spread
Functions (PSFs) relies on the algorithm of \citet{Ala_Lup98}.  Next, in
the {\sc iraf} environment\footnote[10]{{\sc iraf} is distributed by the
National Optical Astronomy Observatory, which is operated by AURA under
cooperative agreement with the National Science Foundation.} we derived the
mean color terms of APO's Spicam from observations of Landolt standards
made in the spring of 1999 and 2000.  Finally, we used the PSF magnitudes
of the SNe and their respective field stars along with the known color
terms to derive the standardized $BVRI$ magnitudes of the SNe.

A similar algorithm was used to carry out image subtraction for SN~1999cl
and a revision of the photometry published by \citet{Kri_etal00}.  This
involved, for insurance, the inclusion of artificial stars of known
brightness in the data images.  We also revised the photometry of some of the
field stars.  The stars in question are numbers 7 through 12 in Table 3 of
\citet{Kri_etal00}.  We found that we needed to add 0.032, 0.025, 0.072,
and 0.062 mag to these $BVRI$ magnitudes, respectively, to match the values
used by \citet{Jha02} and \citet{Jha_etal05a}. These brighter stars were
used for the reduction of the photometry obtained with the MRO 0.9-m
telescope.  Our MRO observations were made at high airmass and with poor
guiding. Jha's photometry of field stars 7 through 12 is much more likely
to be correct.

In Tables \ref{99ccphot}, \ref{99clphot}, and \ref{00cfphot} we present our
$BVRI$ photometry of SNe~1999cc, 1999cl, and 2000cf.

Figures \ref{99cc_fits}, \ref{99cl_lc}, and \ref{00cf_fits} show our $BVRI$
photometry of SNe~1999cc, 1999cl, and 2000cf, along with data of
\citet{Jha_etal05a}.  As one can see, the internal agreement of the SN~1999cc
and SN~1999cl photometry is excellent, indicating that the $BVRI$ filters
used at APO, MRO, and FLWO must have reasonably similar profiles.  The
SN~2000cf photometry agrees quite well in the $B$- and $V$-bands, but the
$R$- and $I$-band photometry shows some systematic differences at $\sim$10 to
20 days after T($B_{max}$).  As Jha (2005, private communication) points out,
the MLCS code accounts for the fact that there is considerable scatter of the
$R$- and $I$-band light curves at this epoch, even for unreddened objects of
identical decline rate observed with only one telescope.  Spectroscopic
corrections \citep{Str_etal02, Kri_etal03} are required for eggregious
differences in data sets comprised of photometry from different telescopes.

We have only one new infrared datum to report.  For SN~2000cf we
derive $H$ = 17.830 $\pm$ 0.052 from a mosaic representing 485 sec of total
integration time and obtained with the APO 3.5-m telescope and Grim II on
11.27 May 2000 UT (= JD 2,451,675.77). This measurement was calibrated by
means of PSF magnitudes of the SN and
the infrared standard stars AS~27-0 and AS~27-1 of
\citet{Hun_etal98}, which were imaged before and after the SN field.  As a
check, we derived the $H$-band magnitudes of three field stars near the SN
and compared them with values from the Two Micron All Sky Survey (2MASS).  
The agreement was within 1$\sigma$.  The brightest of the three is our
``star 1'' of the SN~2000cf photometry sequence. We observed $H$ = 14.229
$\pm$ 0.011, while 2MASS gives $H$ = 14.231 $\pm$ 0.045 for this star.

\section{Discussion}  

\subsection{SNe~1999cc and 2000cf in the Optical Bands}

In Table \ref{solutions} we give the light curve solutions of SNe~1999cc
and 2000cf using the \dmm\ method of \citet{Pri_etal05}.

For SN~1999cc we obtain a host galaxy extinction of A$_V$ $\approx$ 3.1
$\times$ 0.055 = 0.171 mag.  For SN~2000cf A$_V$ (host) $\approx$ 0.084
mag.  \citet{Jha_etal05a} obtain host galaxy extinction of A$_V$ = 0.148
$\pm$ 0.088 for SN~1999cc and A$_V$ = 0.194 $\pm$ 0.094 for SN~2000cf.  
The MLCS light curve solutions of \citet{Jha_etal05a} of these two SNe
are also based on the combination of APO and FLWO data, with no
S-corrections applied.  

On a scale of H$_0$ = 72 \kms\ Mpc$^{-1}$ \citep{Fre_etal01}
\citet{Jha_etal05a} obtained a distance modulus for SN~1999cc of 35.63
$\pm$ 0.09 mag, which agrees with our value of 35.64 $\pm$ 0.17 mag.  
For SN~2000cf they obtain $m-M$ = 36.15 $\pm$ 0.10 mag on an H$_0$ = 72
scale, in excellent agreement with our value of 36.21 $\pm$ 0.17 mag.

\subsection{SN~1999cl and Non-Standard Dust}

The \dmm\ light curve fitting code does not have the functionality to solve
for the reddening and distance of a Type Ia SN suffering extremely
non-standard reddening.  Still, with our revised photometry of SN~1999cl and
the data of \citet{Jha_etal05a} we may derive updated parameters of interest
for this object (see Table \ref{99cl_params}).  Some parameters are quite
straightforward to determine:  the observed times of $B$- and $V$-band
maximum light, and the $B$- and $V$-band maxima themselves.  
\citet{Phi_etal99} point out that the observed decline rate of a reddened
Type Ia SN is made {\em slower} as a result of the reddening.  Whereas the
observed decline rate of SN~1999cl is \dmm\ = 1.175 $\pm$ 0.075 mag, the
true decline rate is roughly 1.285 $\pm$ 0.080 mag.  To obtain this
correction we used the Type Ia SN spectral template of \citet{Nug_etal02}
and used the reddening law of \citet[][hereafter CCM89] {Car_etal89}
modified to be consistent with the non-standard reddening discussed below.
We found that correcting the observed value of \dmm\ to the true value
\citep[e.g.,][Eqn. 6]{Phi_etal99} is only weakly dependent on the value of
R$_V$.

From Eqn. 7 of \citet{Phi_etal99} we can estimate that the pseudo-color
(B$_{max} - V_{max}$)$_0$ = $-$0.049 $\pm$ 0.033 for SN~1999cl.  Given its
observed $B$- and $V$-band maxima, we estimate that the total color excess
is E($B-V$)$_{tot}$ = 1.246 $\pm$ 0.070 mag, of which 0.038 mag is due to
dust in our Galaxy \citep{Sch_etal98}.  Thus, E($B-V$)$_{host}$ = 1.208
$\pm$ 0.070 mag.

\citet{Wan_etal05} have delineated another well-behaved color locus for Type
Ia SNe unreddened in their hosts: the $B-V$ color 12 days after
T($B_{max}$).  We derive a corresponding value of E($B-V$)$_{host}$ = 1.264
$\pm$ 0.070 for SN~1999cl.  Averaging this with the value given above, but
taking a conservative estimate of the uncertainty, we shall adopt
E($B-V$)$_{host}$ = 1.236 $\pm$ 0.070 mag.

Using the method articulated by \citet{Kri_etal00},\footnote[11]{Our
analysis of the $V$ minus near-IR colors of mid-range decliners relied on
SNe 1972E, 1980N, 1983R, and 1999cp to set the zero points of the colors,
with extra information on the shape of the loci from SNe 1981B, 1981D,
1998bu, and 1999cl.  Since we are re-analyzing SN~1999cl here on the basis
of revised optical photometry and with the elimination of two nights of IR
data, we should not use the exact same color loci derived by us previously.  
Here we derive color excesses of SN~1999cl using unreddened loci derived
from the other seven SNe.  Over the past few years we have unfortunately
not obtained well sampled optical and IR light curves of any other
unreddened spectroscopically normal Type Ia SN of mid-range decline rates.}
we derive total color excesses of E($V-J$)$_{tot}$ = 1.552 $\pm$ 0.163,
E($V-H$)$_{tot}$ = 1.793 $\pm$ 0.028, E($V-K$)$_{tot}$ = 1.972 $\pm$ 0.073
mag for SN~1999cl.  To derive the host galaxy reddening and host galaxy
value of R$_V$ we must subtract the contributions due to dust in our
Galaxy. From \citet{Sch_etal98} and standard Galactic dust parameters from
CCM89, we obtain A$_V$(Gal) $\approx$ 3.1 $\times$ 0.038 = 0.118,
E($V-J$)$_{Gal}$ = 0.085, E($V-H$)$_{Gal}$ = 0.095, and E($V-K$)$_{Gal}$ =
0.104 mag.

It is sensible that the color excesses of any highly reddened star or SN
increase monotonically as we increase the baseline of the two filters in
question.  If we consider a fictitious far-infrared photometric band ``X''
with central wavelength $\lambda$,

\begin{equation}
\mathrm{lim}_{\lambda \rightarrow \infty} \; E(V-X_{\lambda}) \; = \; A_V \; \; .
\end{equation}

\parindent = 0 mm

The advantage of using a combination of $V$-band and near-IR data is
that we can scale a color excess by a coefficient not much greater
than 1.0 to obtain A$_V$.  The scaling coefficient is almost the
same whether we have ``normal'' Galactic dust or some more exotic
type.

\parindent = 9 mm

Here we restrict the IR data under consideration to data obtained by us on
four photometric nights when SN~1999cl was calibrated directly by means of
IR standards.  Previously, we reported two other nights of data wherein the
IR photometry of SN~1999cl was calibrated with respect to the core of the
host galaxy.  See Table 7 of \citet{Kri_etal00}. Given that using the core
of a galaxy as a flux reference is seeing dependent, it is probably best to
exclude these other two nights of data from consideration.

In the Appendix we lay out a general means of determining R$_V$, based on
the CCM89 interstellar extinction model, and which is applicable to standard
and non-standard dust.  To derive extinction from observations of SNe,
however, one needs to use CCM89-type coefficients derived from spectra of
SNe, rather than from spectra of normal stars.  We used the generic Type Ia
SN spectrum at T($B_{max}$) of \citet{Nug_etal02} to determine the
coefficients for the $UBVRI$ bands.  Our coefficients listed in Table
\ref{ccm_coeffs} are consistent (i.e., within 0.02) with the values derived
by \citet{Jha02} from 91 spectra of SNe that he measured photometrically.  
For the coefficients corresponding to the $JHK$ bands we used the spectrum
of SN~1999ee at +1 d with respect to T($B_{max}$) of \citet{Ham_etal02}.  
There is a time dependence of these coefficients because of the change in
the spectral energy distributions (SEDs) of SNe.  These time variations have
been quantified by \citet[][Fig. 4.4]{Jha02} for the $UBVRI$ bands.  Within
$\pm$ 10 d of T($B_{max}$) these variations are not large.

For SN~1999cl we determined three estimates of R$_V$.  
From E($V-J$)$_{host}$ we get R$_V$ = 1.36 $\pm$ 0.25.  
From E($V-H$)$_{host}$ we get R$_V$ = 1.53 $\pm$ 0.11.  
From E($V-K$)$_{host}$ we get R$_V$ = 1.63 $\pm$ 0.13.  
The weighted mean is R$_V$ = 1.55 $\pm$ 0.08, and the resulting value of
A$_V$(host) = 1.91 $\pm$ 0.15 mag.  From the CCM89 interstellar
extinction model it follows that A$_{\lambda}$ = [0.161, 
0.102, 0.066] $\times$ A$_V$ for the $J$, $H$, and $K$ bands, 
respectively.

Unfortunately, two sanity checks that we might like to do cannot be done.  
\citet{Lir95} found that the unreddened $B-V$ colors of Type Ia from 30 to
90 days after T$(V_{max}$) delineate a color locus independent of \dmm. We
have no SN~1999cl photometry at this epoch.  Neither do we have an
unreddened $V-I$ locus analogous to our $V-[J,H,K]$ loci. 

A third sanity check involves $I$-band data.  Standard Galactic dust with
$R_V$ = 3.1 would give E($V-I$) = 1.26 $\times$ E($B-V$), using the
coefficients given in Table \ref{ccm_coeffs}.  For $R_V$ = 1.55, E($V-I$) =
0.93 $\times$ E($B-V$).  So we would expect E($V-I$) $\approx$ 1.15 mag.
From our $I$-band light curve we find $V_{max}$ = 13.123 $\pm$ 0.039 at JD
2,451,340.94 $\pm$ 0.50. Using Eqn. 8 of \citet{Phi_etal99} to estimate the
intrinsic pseudo-color $V_{max} - I_{max}$ of SN~1999cl, we obtain
E($V-I$)$_{host}$ = 0.95 $\pm$ 0.06, which is 0.2 mag less than we would
have expected.  \citet[][Fig. 5]{Rei_etal05} have already hinted that
the $B-V$ and $V-I$ color excesses of SN~1999cl do not have a normal ratio.

Our value of R$_V$ for the host galaxy reddening is equal to
half the standard Galactic value, confirming that M~88,
the host of SN~1999cl, has very non-standard dust.
\citet{Wei_Dra01} discuss various recipes for
interstellar dust that have R$_V$ ranging from 2.1 to 5.5, comparable to
the range of values seen in dusty regions of our Galaxy.  In order to
produce dust with very low values of R$_V$ one needs a mix of silicate and
carbonaceous grains with a relatively high percentage of small particles
(10$^{-3} \; \mu$m). While very large grains give grey extinction with
R$_V$ $\rightarrow \infty$, Rayleigh scattering (A$_{\lambda} \propto
\lambda^{-4}$) would give R$_V$ = 1.2 \citep{Dra03}.  

Lensing galaxies have exhibited some very non-standard reddening
values.  \citet{Fal_etal99} find R$_V$ = 1.47 $\pm$ 0.15 and R$_V$ =
7.20 $\pm$ 0.08 for two galaxies with redshifts of 0.96 and 0.68,
respectively.  \citet{Tof_etal00} find R$_V$ between 1.3 and 2.0 for a
$z$~=~0.44 lensing galaxy.  However, \citet{McG_etal05} point out that
if the dust along multiple sightlines to a lensed quasar has different
properties, then the resulting extinction curve is only a measurement
of the {\em difference} of two extinction curves, which allows for the
possibility of any value of R$_V$, even negative values.  The method
used by \citet{Fal_etal99} and \citet{Tof_etal00} only works if one
of the quasar images in a lensed system is lightly reddened or both
sightlines have the same dust properties.  \citet{McG_etal05} assert
that these cases are likely rare and hard to confirm.  Thus, we must
treat with caution the specific values of R$_V$ obtained with
the method of lensing galaxies.

The latest version of MLCS \citep{Jha_etal05b} has the functionality
to solve for R$_V$ as part of the light curve solutions.  
These authors used a prior distribution on the possible values
of R$_V$ and obtained 2.22 $\pm$ 0.15 for SN~1999cl.  However, Jha
(2005, private communication) indicates that without a prior, he
obtains R$_V <$ 1.8 for SN~1999cl, which is consistent with our
finding here.  A more extensive analysis of reddening toward Type Ia
SNe which have A$_V >$ 0.5, allowing the determination of R$_V$, will
be presented by \citet{Jha_etal06}.

Spectra of SN~1999cl obtained at the Fred L. Whipple Observatory at Mt.  
Hopkins indicate that SN~1999cl was a highly reddened object, but otherwise
spectroscopically normal (Matheson 2005, private communication).  This
allows us to put together a coherent solution for this object.
\citet{Pri_etal05} find that the absolute $V$-band magnitudes at maximum of
Type Ia SNe are related to the decline rates as follows:

\begin {equation}
M_V \; = \; (-19.246 \; \pm \; 0.019) \; + \; (0.606 \; \pm \; 0.069) \; \times \; [\dmm \; - \; 1.1] \; \; , 
\end {equation}

\parindent = 0 mm

where the zeropoint of the absolute magnitudes is associated with a
Hubble constant of 72 \kms\ Mpc$^{-1}$ \citep{Fre_etal01}. The RMS
uncertainty of the $V$-band absolute magnitudes is $\pm$~0.14 mag.  With
an absolute magnitude plus apparent magnitude at maximum, and a $V$-band
extinction derived from a combination of $BV$ and near-IR data, we
obtain a distance modulus for SN~1999cl and its host (M~88) of $m-M$ =
30.96 $\pm$ 0.20 mag.  This corresponds to a distance of
15.5$^{+1.5}_{-1.4}$ Mpc, consistent with M~88 being in the
Virgo cluster.  \citet{Ton_etal00} give a mean distance modulus of $m-M$
= 31.03 $\pm$ 0.06 mag for the Virgo cluster, based on the method of
surface brightness fluctuations (SBFs).  Their Virgo distance modulus
is on a scale of H$_0$ = 77 \kms Mpc$^{-1}$.  On an H$_0$ = 72 scale
this corresponds to a distance of 17.2 Mpc.

\parindent = 9 mm

The actual radial velocity of M~88 (= NGC 4501) is 2281 \kms\ from NED.
The Hubble flow velocity, given the distance quoted above, is 1120 \kms.  
Qualitatively speaking, the photometry of SN~1999cl allows us to conclude
that M~88 is on the near side of the Virgo cluster and falling into it at a
relative velocity greater than 1100 \kms.  It is not a surprise to
confirm that the Virgo cluster is a giant gravitational well.

Finally, let us compare the extinction-corrected $V$-band magnitude at
maximum of SN~1999cl with values for other spectroscopically normal Type Ia
SNe in the Virgo cluster. \citet{Rei_etal05} make this easy.  Their Table 1
indicates that the SNe to consider are 1960F, 1981B, 1984A, 1990N, and
1994D.  In their Table 2 they give extinction-corrected $V$-band magnitudes
at maximum of 11.171, 11.827, 11.693, 12.560, and 11.978, respectively.  
Photometry of SN~1960F was photographic, so of lesser quality.  We do not
understand why SN~1990N was so much fainter.  The average of the other three
is $V_{max}^0$ = 11.83 $\pm$ 0.08.  Our corresponding value for SN~1999cl is
$V_{max}^0$ = 11.83 $\pm$ 0.15. Thus, the brightness of SN~1999cl is
entirely consistent with its host, M~88, being in the Virgo cluster.

While all of the individual parameters given in Table
\ref{99cl_params} should be regarded with some degree of caution, we
have a coherent set of parameters for this object.  The most
interesting parameter is the very small value of R$_V$.  Clearly, it
is unwise to assume that dust in other galaxies is always like the
average dust in our Galaxy, with R$_V$ = 3.1

\subsection{The Standard Candle Nature of Type Ia SNe in the 
Near-Infrared}

\citet{Phi_etal99} showed that Type Ia SNe are standardizable candles in
the optical $BVI$ bands.  They presented quadratic relationships that
showed how the absolute magnitudes at maximum varied as a function of the
decline rate parameter \dmm.  The slopes of these decline rate relations
decrease as one proceeds to longer wavelength photometric bands.

\citet{Mei00} first suggested that Type Ia SNe at roughly two weeks after
T($B_{max}$) may be standard candles in the near-infrared, not just
standardizable candles.  \citet[][Fig. 13]{Kri_etal03} showed updated
$BVI$ decline rate relations and also plotted the $H$-band absolute
magnitudes 10 days after T($B_{max}$).  We confirmed that the slope of
the $H$-band decline rate relation was very shallow for objects with
\dmm\ $\leq$ 1.8. \citet{Kri_etal04a} and \citet{Kri_etal04b,
Kri_etal04c} showed that one could construct uniform $JHK$ light curve
templates within 10 days of the time of T($B_{max}$) that allowed one to
estimate accurately what the maximum magnitudes of Type Ia SNe are,
providing that one has one or more nights of observations in the $\pm$10
d time window.  Indeed, the evidence shows that Type Ia SNe appear to be
standard candles in the near-IR over a wide range of \dmm, not just
standardizable candles.

Using our $H$-band light curve template \citep[][Fig. 9 and Table
12]{Kri_etal04b}, a K-correction of +0.048 mag applied to our datum at
3.8 days after the time of $B$-band maximum \citep[][Table
11]{Kri_etal04b}, and an extinction correction of $A_H$ = 0.035 mag, we
estimate that $H$ = 17.59 $\pm$ 0.10 for the K-corrected,
extinction-corrected apparent magnitude of SN~ 2000cf at maximum light.  
With a redshift in the CMB frame of 10803 \kms\ and a Hubble constant of
72 \kms\ Mpc$^{-1}$ \citep{Fre_etal01}, we obtain an absolute magnitude
of $M_H$ = $-$18.29 $\pm$ 0.11. (We have included the effect of a
peculiar velocity of $\pm$~300 \kms\ in the uncertainty of the absolute
magnitude.)  This compares very well with the mean absolute $H$-band
magnitude of 21 other Type Ia SNe with \dmm\ $\leq$ 1.8 \citep[][Table
17]{Kri_etal04c}, namely $\langle M_H$(max)$\rangle$ = $-$18.28 $\pm$
0.03.

To derive the extinction-corrected near-IR absolute magnitudes at
maximum of SN~1999cl let us assume the distance modulus  
derived above by means of $BV$ photometry and $V$ minus near-IR 
colors. Using the scaling factors A$_{\lambda}$/A$_V$ appropriate
for R$_V$ = 1.55, we find that A$_J$ (tot) = 0.34, A$_H$ (tot) = 0.22, and 
A$_K$ (tot) = 0.14 mag, with uncertainties of $\pm$ 0.06.
We obtain the following absolute magnitudes for SN~1999cl at maximum
light: M$_J$ = $-$18.50, M$_H$ = $-$18.20, and M$_K$ = $-$18.52, with 
uncertainties of $\pm$ 0.21 mag. These are to be compared with
the mean values given in Table 17 of \citet{Kri_etal04c}, namely 
$\langle M_J \rangle$ = $-$18.61,
$\langle M_H \rangle$ = $-$18.28,
$\langle M_K \rangle$ = $-$18.44, 
with 1$\sigma$ uncertainties of the distributions of values of $\pm$ 0.13,
0.15, and 0.14 mag, respectively. Taken at face value, the near-IR absolute
magnitudes of SN~1999cl are within 1$\sigma$ of what we would expect.
But how close to ``normal'' the infrared absolute
magnitudes are depends on what the true $V$-band absolute magnitude
is, and we {\em assumed} a value of M$_V$ exactly equal to the mean value
for Type Ia SNe with the decline rate of SN~1999cl.  Without a
measure of the distance to M~88 via Cepheids or the planetary
nebula luminosity function (PNLF), it is probably best not to include 
SN~1999cl in a statistical analysis of the mean absolute magnitudes of Type 
Ia SNe.  While the SBF method is usually limited to elliptical galaxies,
it might be possible to obtain an SBF distance to the central bulge of
M~88.

\section{Conclusions}

We have presented previously unpublished $BVRI$ photometry of SNe~1999cc
and 2000cf.  We have combined our photometry with that of \citet{Jha_etal05a}
and have derived the host galaxy reddening and distances to these objects. 
 
$BVRI$ photometry of SN~1999cl previously published by us
\citep{Kri_etal00} needed
revision for two reasons: 1) different photometry for the field stars
used to calibrate the imagery obtained at Manastash Ridge Observatory; 
and 2) image subtraction templates were obtained in February of 2000.
We have combined our previously published IR photometry with our
revised optical photmetry and the photometry of \citet{Jha_etal05a}
to determine the distance and unusual reddening of this object.
We find that SN~1999cl was reddened by highly non-standard
dust, with R$_V$ = 1.55 $\pm$ 0.08.  This is one of 
the lowest known values of R$_V$, and implies a large
fraction of very small (10$^{-3} \mu$m) grains in the vicinity of
SN~1999cl in its host galaxy.

SNe~1999cl and 2000cf have infrared absolute magnitudes at maximum in
close agreement with the mean of other Type Ia SNe studied by us
and which have \dmm\ $\leq$ 1.8 \citep{Kri_etal04a, Kri_etal04b, 
Kri_etal04c}.

\acknowledgments

We thank Brian Skiff for determining the coordinates of the field stars of
SNe~1999cc and 2000cf. We made use of the NASA/IPAC Extragalactic Database
(NED), data of the Two Micron All Sky Survey, and {\sc simbad}, operated at
CDS, Strasbourg, France.  We thank Bruce Draine for useful discussions and
pointing us to key references; John Tonry; and Saurabh Jha, for useful
discussions and for making data available to us ahead of publication.  K.K.
is partly funded through STScI grant AR-9925 and NSF grants AST-0206329 and
AST-0443378 (the ESSENCE Project).  J.-L.G.R. was supported by the REU
program of the National Science Foundation.  J.L.P. is supported by an OSU
Astronomy Department Fellowship.

\appendix

\section{General Determination of R$_V$}

The universal reddening law of CCM89 is parametrized by 
R$_{V}$ in the form

\begin{equation}
\label{eq:A1}
\frac{A_{\lambda}}{A_{V}} \; = \; a_{\lambda} \; + \; \frac{b_{\lambda}}{R_{V}} \; \; ,
\end{equation}

\parindent = 0 mm

where A$_V$ = R$_V \times$ E($B-V$). We can then rewrite the
previous equation as

\begin{equation}
\label{eq:A2}
A_{\lambda} \; = \; E(B-V)\; \times \; (R_{V} \; a_{\lambda} \; + \; b_{\lambda}) \; \; .
\end{equation}

Now, by definition for some filter ``X'' with central wavelength $\lambda$

\begin{equation}
\label{eq:A3}
E(V-X_{\lambda}) \; = \; A_{V} \; - \; A_{\lambda} \; \; . 
\end{equation}

Equations (\ref{eq:A2}) and (\ref{eq:A3}) combined give us

\begin{equation}
\label{eq:A4}
E(V-X_{\lambda}) \; = \; E(B-V)\; \times \; (R_{V} \; - \; R_{V}a_{\lambda} \; - \; b_{\lambda}) \; \; .
\end{equation}

Finally, we can obtain R$_V$ in terms of the color excesses E($B-V$) 
and E($V-$X$_{\lambda}$) given values of $a_{\lambda}$ and $b_{\lambda}$:

\begin{equation}
\label{eq:A5}
R_V \; = \; \frac{1}{1-a_{\lambda}}\left [ \frac{E(V-X_{\lambda})}{E(B-V)} \; + \; b_{\lambda} \right ] \; \; .
\end{equation}

Propagating the errors in the color excesses, and ignoring any uncertainty in
$a_{\lambda}$ and $b_{\lambda}$ , we obtain

\begin{equation}
\label{eq:A6}
\sigma_{R_{V}} \; = \; \frac{1}{|1-a_{\lambda}|}\times \frac{E(V-X_{\lambda})}
{E(B-V)} \times \sqrt{ \left( \frac{\sigma_{E(B-V)}}{E(B-V)} \right)^{2} \; + \; 
\left ( \frac{\sigma_{E(V-X_{\lambda})}}{E(V-X_{\lambda})} \right )^{2} } \; \; .
\end{equation}

The values of the color excesses in all the previous equations assume that we have
a unique value of R$_V$.  To determine the host galaxy value of R$_V$ one should
first subtract off the contributions of Galactic reddening to the total color
excesses.  We note also that Eqn. A6 relates only to the random errors of the
color excesses.  Systematic errors are not taken into account, such as incorrectly
assuming what the unreddened colors of any particular SN are, which of course
would lead to systematic errors in the derived color excesses.

\parindent = 9 mm

An update to the CCM89 model can be found in \citet{Val_etal04}.  Any
differences with CCM89, for example in the case of galaxies with LMC- or
SMC-like reddening laws with different metallicity, are most relevant for
far-UV reddening \citep{Gor_etal03}.  For our purposes here the
parameterization and coefficients of CCM89 are useful but should be modified
slightly. If we attempt to derive interstellar extinction based on broad
band photometry of SNe, we need to use coefficients of the reddening model
based on the SED of the SNe themselves, not the SEDs of normal stars.
\citet[][\S4.2.2, Fig. 4.5, and Table 4.1]{Jha02} used a sample of 91
spectra of Type Ia SNe observed by him photometrically.  We confirmed his
coefficients (within 0.02) for the $UBVRI$ bands using the generic Type Ia
SN spectral template of \citet{Nug_etal02}.  For the near-IR bands we used a
spectrum of SN~1999ee at +1 d with respect to T($B_{max}$) from
\citet{Ham_etal02}, first dereddened, then reddened according to various
values of R$_V$.  Our coefficients for optical and near-IR bands are given
in Table \ref{ccm_coeffs} and are most appropriate for photometry obtained
near the time of maximum light.  The IR filter profiles used correspond to
the filters used at the Las Campanas 1-m telescope for setting up the
standard star system of \citet{Per_etal98}.

\parindent = 0 mm

{\bf Note added for this preprint}: \citet{Wan05} suggests that very low values
of R$_V$ may result from the scattering of the light of a SN by dust
clouds in the circumstellar environment.  The scattered light reduces the
effective R$_{\lambda}$ in the optical, but has the opposite effect in
the ultraviolet.

\begin{deluxetable}{ccccccc}
\tablewidth{0pt}
\tablecolumns{7}
\tablecaption{NGC 6038 Photometric Sequence for SN 1999cc\label{99cccoords}}
\tablehead{
\colhead{$\star$} & \colhead{$\alpha$ (2000)$^a$} &
\colhead{$\delta$ (2000)$^a$} & \colhead {$B$} & \colhead {$V$} &
\colhead{$R$} & \colhead{$I$} }
\startdata
SN & 16:02:42.0 & +37:21:34 &                 &                &                &                \\
 2 & 16:02:44.8 & +37:19:37 &  17.301 (0.018) &  16.371 (0.014) &  15.836 (0.017) &  15.330 (0.015) \\
 3 & 16:02:34.5 & +37:20:36 &  16.221 (0.017) &  15.648 (0.011) &  15.306 (0.016) &  14.931 (0.013) \\
 4 & 16:02:48.6 & +37:21:03 &  20.161 (0.019) &  18.748 (0.011) &  17.858 (0.014) &  16.978 (0.014) \\
 5 & 16:02:51.1 & +37:21:12 &  19.534 (0.017) &  19.175 (0.010) &  18.941 (0.016) &  18.695 (0.019) \\
 6 & 16:02:44.2 & +37:21:50 &  18.666 (0.015) &  18.033 (0.010) &  17.655 (0.015) &  17.244 (0.013) \\
 7 & 16:02:43.0 & +37:21:58 &  19.415 (0.014) &  18.072 (0.011) &  17.252 (0.018) &  16.488 (0.013) \\
 8 & 16:02:41.2 & +37:22:51 &  19.558 (0.018) &  18.983 (0.012) &  18.628 (0.016) &  18.251 (0.017) \\
 9 & 16:02:39.2 & +37:23:21 &  17.706 (0.016) &  17.154 (0.010) &  16.821 (0.016) &  16.451 (0.012) \\
10 & 16:02:32.9 & +37:23:17 &  19.508 (0.024) &  18.555 (0.006) &  17.993 (0.008) &  17.511 (0.012) \\
\enddata
\tablenotetext{a} {The coordinates of stars 5, 6, and 7 were
estimated from Digital Sky Survey images using SkyView.  Coordinates of all other
field stars were determined from the USNO-A2.0 catalog.}
\end{deluxetable}

\begin{deluxetable}{ccccccc}
\tablewidth{0pt}
\tablecolumns{7}
\tablecaption{MCG +11$-$19$-$25 Photometric Sequence for SN 2000cf\label{00cfcoords}}
\tablehead{
\colhead{$\star$} & \colhead{$\alpha$ (2000)$^a$} &
\colhead{$\delta$ (2000)$^a$} & \colhead {$B$} & \colhead {$V$} &
\colhead{$R$} & \colhead{$I$} }
\startdata
SN &  15:52:56.2 & +65:56:13 &             &                &                &                \\
 1 &  15:52:58.0 & +65:55:36 & 16.672 (0.024) &  15.963 (0.005) &  15.539 (0.010) &  15.146 (0.017) \\
 2 &  15:52:54.3 & +65:56:56 & 18.654 (0.022) &  17.897 (0.009) &  17.425 (0.014) &  16.963 (0.017) \\
 3 &  15:52:41.7 & +65:57:29 & 17.507 (0.019) &  16.875 (0.006) &  16.470 (0.009) &  16.101 (0.017) \\
 4 &  15:52:36.1 & +65:57:08 & 19.166 (0.039) &  18.126 (0.017) &  17.492 (0.021) &  16.888 (0.025) \\
 5 &  15:52:47.9 & +65:55:09 & 18.481 (0.028) &  17.786 (0.006) &  17.347 (0.012) &  16.940 (0.019) \\
 6 &  15:52:42.6 & +65:54:46 & 18.485 (0.025) &  17.090 (0.006) &  16.234 (0.012) &  15.466 (0.017) \\
 7 &  15:53:08.8 & +65:55:55 & 20.993 (0.073) &  19.457 (0.029) &  18.309 (0.035) &  16.902 (0.045) \\

\enddata
\tablenotetext{a} {Field star coordinates were determined from the USNO-A2.0 catalog, except
star 7, whose position was determined from a SkyView DSS image.}
\end{deluxetable}

\begin{deluxetable}{lcccc}
\tablewidth{0pt}
\tablecolumns{5}
\tablecaption{Photometry of SN~1999cc\label{99ccphot}}
\tablehead{
\colhead{JD\tablenotemark{a}} &
\colhead{$B$} &
\colhead{$V$} &
\colhead{$R$} &
\colhead{$I$}
}
\startdata
313.8365 &  16.880 (0.032) &  16.870 (0.017) &  \nodata        &  16.968 (0.027) \\
314.8194 &  16.826 (0.024) &  16.829 (0.011) &  16.737 (0.014) &  16.965 (0.021) \\
314.8293 &  16.866 (0.014) &  16.825 (0.013) &  16.730 (0.015) &  16.953 (0.016) \\
315.6554 &  16.875 (0.016) &  16.809 (0.021) &  16.707 (0.027) &  16.995 (0.023) \\
315.6648 &  16.863 (0.016) &  16.820 (0.016) &  16.712 (0.018) &  17.001 (0.022) \\
316.7991 &  16.893 (0.013) &  16.811 (0.017) &  16.711 (0.013) &  17.042 (0.011) \\
316.8117 &  16.898 (0.011) &  16.815 (0.010) &  16.711 (0.011) &  17.027 (0.013) \\
317.8244 &  16.923 (0.013) &  16.798 (0.016) &  16.691 (0.015) &  17.038 (0.012) \\
317.8356 &  16.910 (0.024) &  16.831 (0.010) &  \nodata        &  17.048 (0.017) \\
318.6731 &  16.990 (0.021) &  16.825 (0.012) &  \nodata        &  17.071 (0.013) \\
322.6629 &  \nodata        &  16.954 (0.048) &  \nodata        &  \nodata        \\
336.6431 &  19.086 (0.017) &  17.918 (0.013) &  17.561 (0.026) &  17.438 (0.021) \\
336.6700 &  19.063 (0.024) &  17.896 (0.009) &  17.526 (0.014) &  17.440 (0.013) \\
339.8234 &  19.357 (0.029) &  18.136 (0.020) &  \nodata        &  17.453 (0.014) \\
\enddata
\tablenotetext{a} {Julian Date minus 2,451,000.}
\end{deluxetable}

\begin{deluxetable}{lccccc}
\tablewidth{0pt}
\tablecolumns{6}
\tablecaption{Revised Photometry of SN~1999cl\label{99clphot}}
\tablehead{
\colhead{JD\tablenotemark{a}} &
\colhead{Observatory} &
\colhead{$B$} &
\colhead{$V$} &
\colhead{$R$} &
\colhead{$I$}
}
\startdata
336.6966   & APO & 15.287 (0.013) & 14.215 (0.007) & 13.585 (0.014) & 13.219 (0.016) \\
337.7858   & APO & 15.213 (0.024) & 14.138 (0.014) & 13.578 (0.027) & 13.265 (0.029) \\
340.6926   & APO & 15.032 (0.011) & 13.902 (0.006) & 13.391 (0.011) & 13.130 (0.012) \\
339.7589   & MRO & 15.124 (0.024) & 13.957 (0.008) & 13.404 (0.018) & 13.075 (0.021) \\
363.7404   & MRO & 16.873 (0.072) & 14.781 (0.010) & 14.066 (0.018) & 13.391 (0.019) \\
365.7448   & MRO & 16.839 (0.038) & 14.904 (0.008) & 14.141 (0.017) & 13.403 (0.019) \\
367.7305   & MRO & 16.886 (0.178) & 14.945 (0.018) & 14.226 (0.031) & 13.363 (0.030) \\
368.7221   & MRO & 17.050 (0.107) & 15.113 (0.023) & 14.244 (0.040) & 13.461 (0.039) \\
\enddata
\tablenotetext{a} {Julian Date minus 2,451,000.}
\end{deluxetable}

\begin{deluxetable}{lcccc}
\tablewidth{0pt}
\tablecolumns{5}
\tablecaption{Photometry of SN~2000cf\label{00cfphot}}
\tablehead{
\colhead{JD\tablenotemark{a}} &
\colhead{$B$} &
\colhead{$V$} &
\colhead{$R$} &
\colhead{$I$}
}
\startdata
675.6850 &  17.222 (0.021) &  17.201 (0.026) &  17.096 (0.013) &  17.398 (0.021) \\
684.6655 &  18.023 (0.021) &  17.577 (0.009) &  17.571 (0.014) &  17.901 (0.021) \\
686.6620 &  18.219 (0.018) &  17.688 (0.016) &  17.643 (0.014) &  17.921 (0.023) \\
689.6957 &  18.611 (0.042) &  17.883 (0.017) &  17.760 (0.022) &  17.858 (0.034) \\
693.6902 &  19.119 (0.018) &  18.140 (0.017) &  17.844 (0.013) &  17.843 (0.015) \\
707.6905 &  20.171 (0.040) &  18.962 (0.022) &  18.412 (0.018) &  18.114 (0.026) \\
711.6762 &  20.269 (0.048) &  19.133 (0.023) &  18.671 (0.018) &  18.369 (0.019) \\
\enddata
\tablenotetext{a} {Julian Date minus 2,451,000.}
\end{deluxetable}

\begin{deluxetable}{lrr}
\tablewidth{0pt}
\tablecolumns{3}
\tablecaption{Light Curve Solutions\label{solutions}}
\tablehead{
\colhead{Parameter/Object:} &
\colhead{SN~1999cc} &
\colhead{SN~2000cf} 
}
\startdata
Redshift\tablenotemark{a}     & 9452 \kms     & 10803 \kms    \\
JD of T($B_{max}$)            & 2,451,315.99 (0.09) & 2,451,671.90 (0.28) \\
\dmm\                         & 1.41 (0.02) mag   & 1.10 (0.01) mag  \\
E($B-V$)$_{Gal}$\tablenotemark{b} & 0.023 mag     & 0.032 mag     \\
E($B-V$)$_{host}$             & 0.055 (0.006) mag  & 0.027 (0.013) mag   \\
($m-M$)\tablenotemark{c}      & 35.64 (0.17) mag & 36.21 (0.17) mag   \\
$B_{max}$\tablenotemark{d}    & 16.84 (0.02) & 17.15 (0.04) \\
$V_{max}$\tablenotemark{d}    & 16.81 (0.02) & 17.14 (0.04) \\
$R_{max}$\tablenotemark{d}    & 16.69 (0.02) & 17.03 (0.04) \\
$I_{max}$\tablenotemark{d}    & 16.93 (0.02) & 17.30 (0.04) \\
$\chi^2_{\nu}$                & 0.79          & 2.04          \\
N\tablenotemark{e}            & 65            & 61            \\
\enddata
\tablenotetext{a}{From NASA/IPAC Extragalactic Database (NED).  Host redshift coverted
to frame of Cosmic Microwave Background.}
\tablenotetext{b}{From \citet{Sch_etal98}.}
\tablenotetext{c}{Distance modulus on a scale of H$_0$ = 72 \kms\  Mpc$^{-1}$.
We include the scatter of the nearby training set of SNe.}
\tablenotetext{d}{These maximum apparent magnitudes do {\em not} include any
K-corrections or extinction corrections.  They are the observed maxima.}
\tablenotetext{e}{Number of data points, including data of 
\citet{Jha_etal05a}.}
\end{deluxetable}

\begin{deluxetable}{lr}
\tablewidth{0pt}
\tablecolumns{2}
\tablecaption{Derived Parameters for SN~1999cl\label{99cl_params}}
\tablehead{
\colhead{Parameter} &
\colhead{Value}
}
\startdata
JD of T($B_{max}$)            & 2,451,341.76 $\pm$ 0.29  \\
JD of T($V_{max}$)            & 2,451,343.96 $\pm$ 0.50 \\
$B_{max}$                     & 15.054 (0.056)  \\
$V_{max}$                     & 13.857 (0.027)  \\
\dmm\ (obs)                   & 1.175 (0.075)     \\
\dmm\ (true)                  & 1.285 (0.080)    \\
($B_{max} - V_{max}$)$_0$\tablenotemark{a}     & $-$0.049 (0.033) \\
E($B-V$)$_{Gal}$\tablenotemark{b}     & 0.038 (0.004)    \\
E($B-V$)$_{host}$             & 1.236 (0.070)    \\
E($V-J$)$_{tot}$              & 1.552 (0.163)    \\
E($V-H$)$_{tot}$              & 1.793 (0.028)    \\
E($V-K$)$_{tot}$              & 1.972 (0.073)    \\
R$_V$(host)                   & 1.55 (0.08)      \\
A$_V$(host)                   & 1.91 (0.15)    \\
A$_V$(tot)                    & 2.03 (0.15)    \\
M$_V$\tablenotemark{c}        & $-$19.13 (0.14) \\
$m-M$                         & 30.96 (0.20) mag  \\
Distance                      & 15.5$^{+1.5}_{-1.4}$ Mpc \\
\enddata
\tablenotetext{a}{From \citet{Phi_etal99}, Eqn. 7.}
\tablenotetext{b}{From \citet{Sch_etal98}.}
\tablenotetext{c}{Given true decline rate and assuming
decline rate relation of \citet{Pri_etal05}.}
\end{deluxetable}

\begin{deluxetable}{cccc}
\tablewidth{0pt}
\tablecolumns{4}
\tablecaption{Coefficients and Extinction in Standard Filters\tablenotemark{a}\label{ccm_coeffs}}
\tablehead{
\colhead{Filter} &
\colhead{$a_{\lambda}$} &
\colhead{$b_{\lambda}$} &
\colhead{$a_{\lambda} + b_{\lambda}$/3.1} 
}
\startdata
$U$   &  0.9685  &  \phs1.7369  &   1.5288 \\
$B$   &  1.0113  &  \phs0.9406  &   1.3147 \\
$V$   &  1.0000  &  \phs0.0000  &   1.0000 \\
$R$   &  0.9217  &   $-$0.2846  &   0.8299 \\
$I$   &  0.7826  &   $-$0.5506  &   0.6050 \\
$J$   &  0.4042  &   $-$0.3764  &   0.2828 \\
$H$   &  0.2550  &   $-$0.2374  &   0.1785 \\
$K$   &  0.1648  &   $-$0.1532  &   0.1154 \\
\enddata
\tablenotetext{a} {Coefficients for calculating extinction for different 
values of R$_V$, as in the reddening model of \citet{Car_etal89}, but 
based on spectra of Type Ia SNe.  Technically, these coefficients
are only valid within several days of the time of maximum light.  The last 
column gives the ratio A$_{\lambda}$/A$_V$ for the case of ``normal'' 
Galactic dust with R$_V$ = 3.1.}
\end{deluxetable}

\clearpage

\figcaption[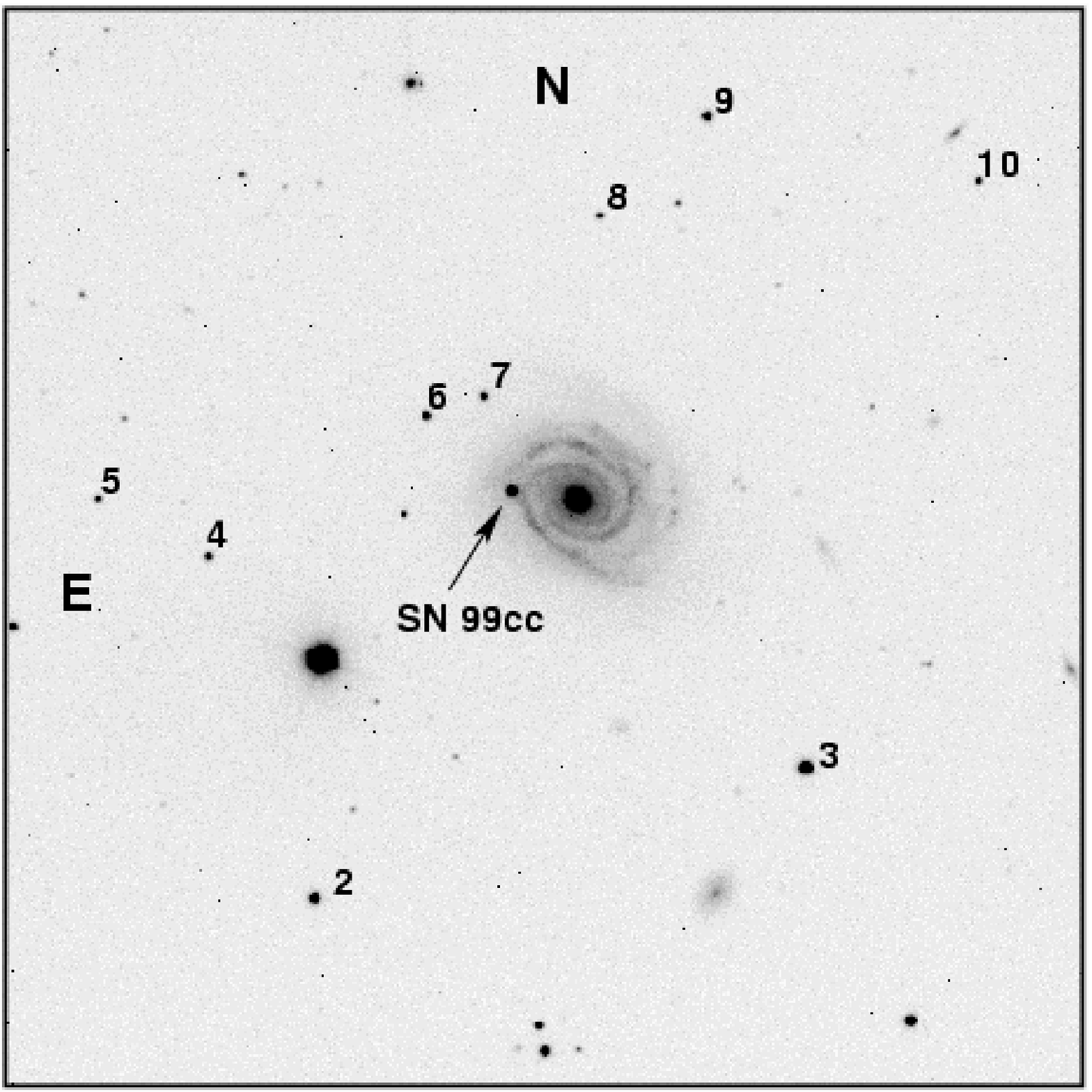] {A $V$-band image of NGC 6038 obtained at APO on
15 May 1999 UT, showing SN~1999cc and the stars of its photometric sequence.
The field is 4.8 arcmin on a side.  North is up, east to the left. \label{99cc_finder}
}

\figcaption[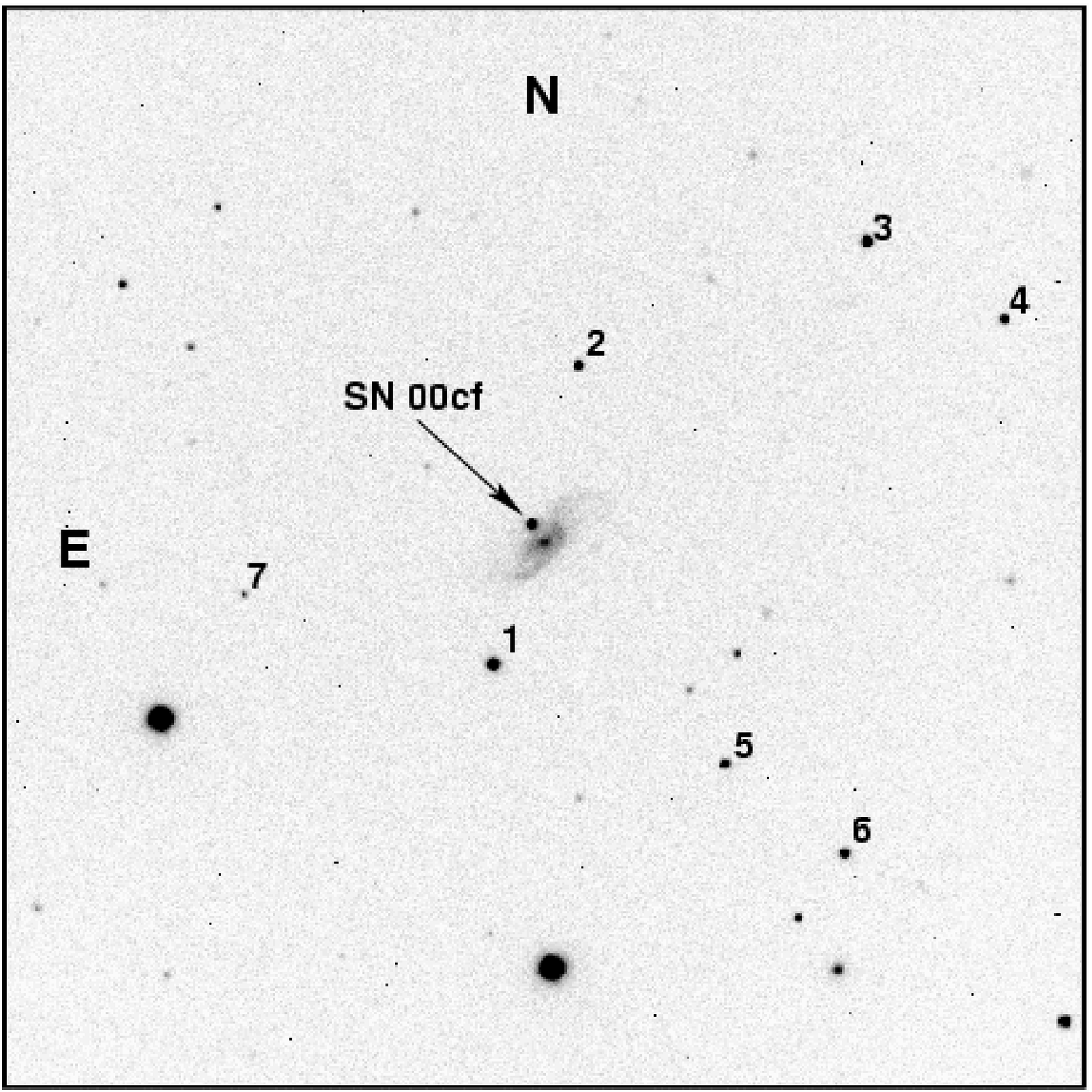] {A $V$-band image of MCG $+11-19-25$ obtained at APO on
11 May 2000 UT, showing SN~2000cf and the stars of its photometric sequence.
The field is 4.8 arcmin on a side.  North is up, east to the left. \label{00cf_finder}
}

\figcaption[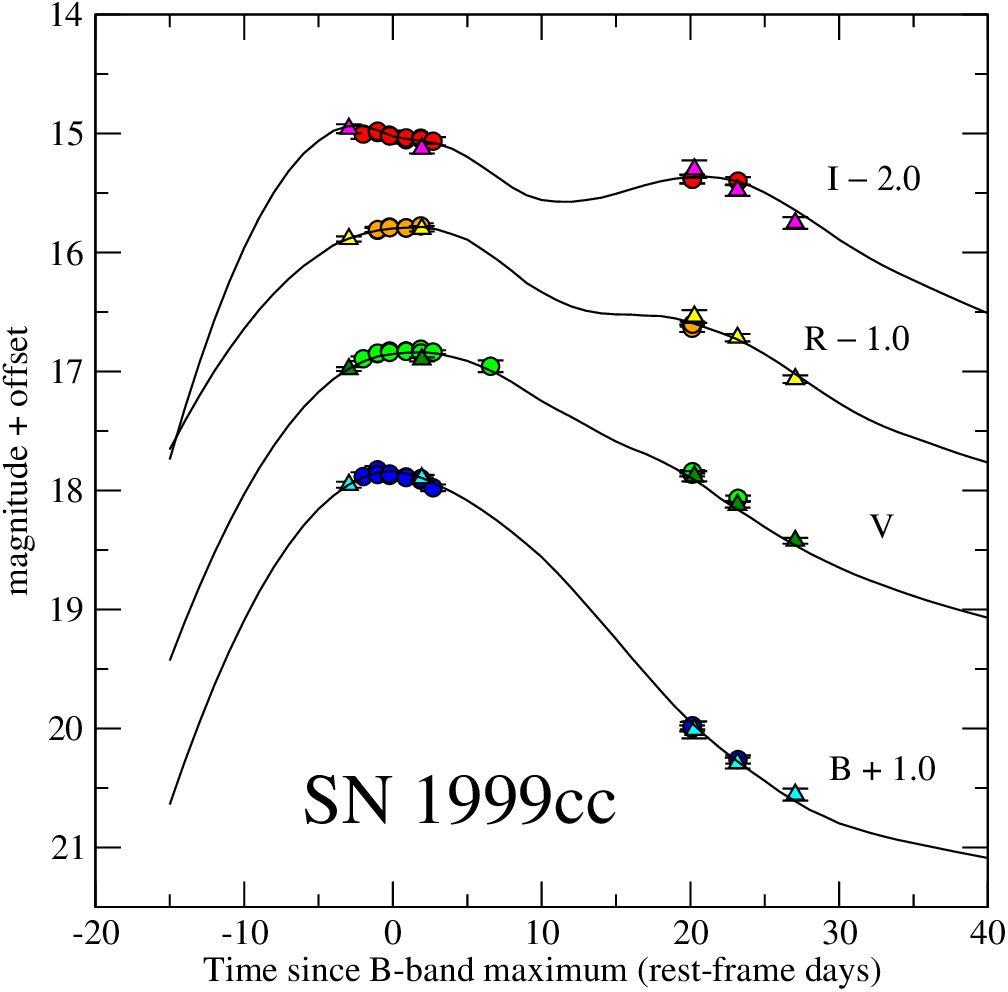] {$BVRI$ light curves of SN~1999cc, showing
our photometry (dots) and that of \citet[][triangles]{Jha_etal05a}.
We fit the $B$- and $V$-band data with third order polynomials for the
determination of the maxima, times of the maxima, and \dmm.
\label{99cc_fits}
}

\figcaption[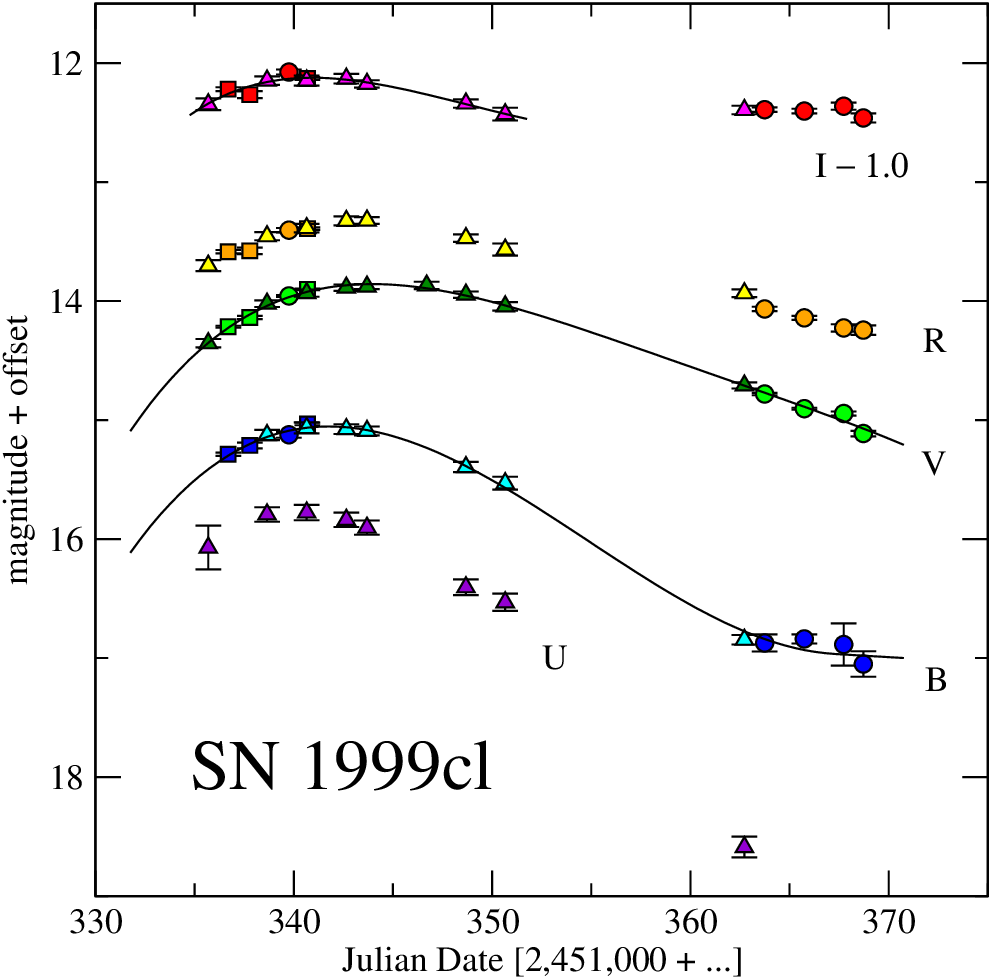] {$UBVRI$ light curves of SN~1999cl, showing
our APO photometry (squares), MRO photometry (dots), and that of 
\citet[][triangles]{Jha_etal05a}.  Notice that except for the $I$-band
data the photometry has been plotted with no offsets, indicating that
this object is extremely reddened.
\label{99cl_lc}
}

\figcaption[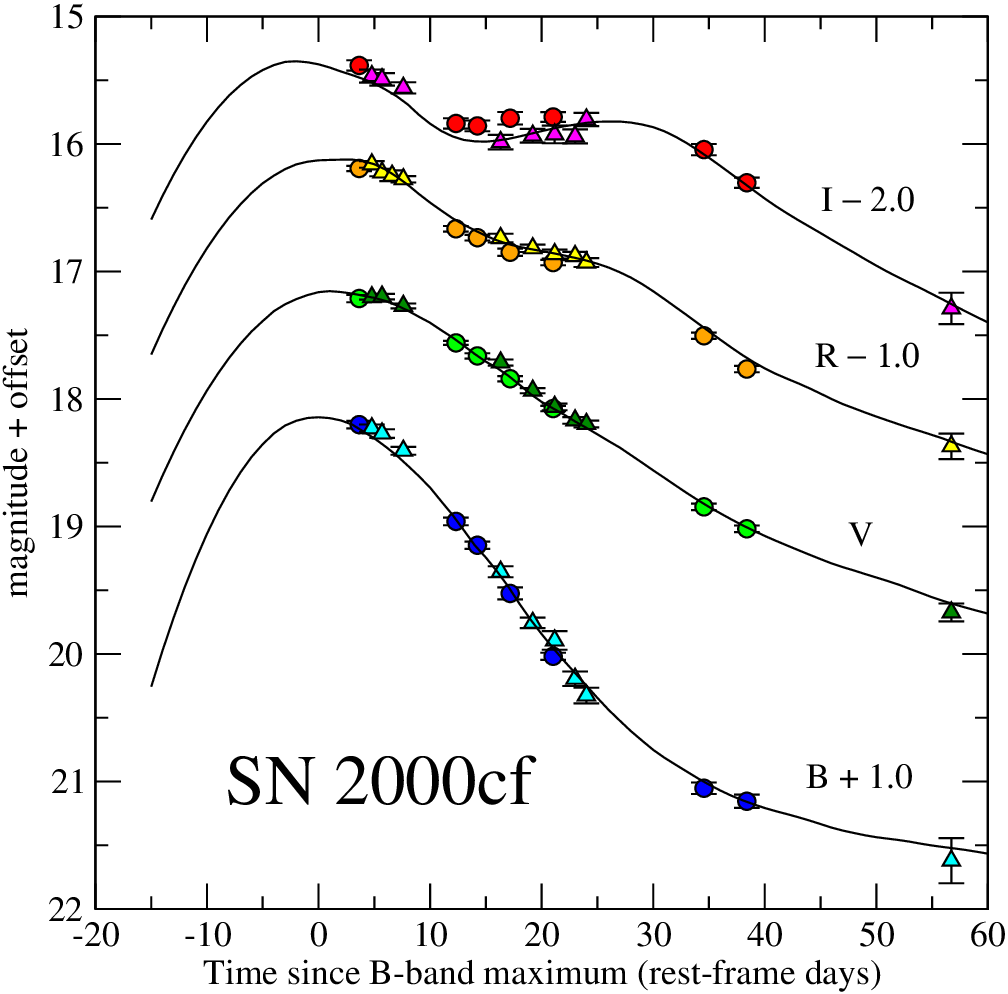] {$BVRI$ light curves of SN~2000cf, showing
our photometry (dots) and that of \citet[][triangles]{Jha_etal05a}.
\label{00cf_fits}
}

\clearpage

\begin{figure}
\plotone{f99cc.ps}
{\center Krisciunas {\it et al.} Fig. \ref{99cc_finder}}
\end{figure}

\begin{figure}
\plotone{f00cf.ps}
{\center Krisciunas {\it et al.} Fig. \ref{00cf_finder}}
\end{figure}

\begin{figure}
\plotone{99cc_fits.eps}
{\center Krisciunas {\it et al.} Fig. \ref{99cc_fits}}
\end{figure}

\begin{figure}
\plotone{99cl_lc.eps}
{\center Krisciunas {\it et al.} Fig. \ref{99cl_lc}}
\end{figure}

\begin{figure}
\plotone{00cf_fits.eps}
{\center Krisciunas {\it et al.} Fig. \ref{00cf_fits}}
\end{figure}

\end{document}